\documentclass[prl,amsmath,amssymb,amsfonts,twocolumn,nofootinbib]{revtex4}
\usepackage{bm,graphicx,mathrsfs}

\usepackage{graphicx}
\usepackage{epsfig}
\usepackage{amsmath,bbm}
\usepackage{amsfonts,amssymb}
\usepackage{times}

\newcommand{\ii}{\mathbbm{1}}
\newcommand{\M}{{\cal M}}

\newcommand{\1}{\mathbbm{1}}
\newcommand{\id}{{\rm id}}

\newcommand{\tr}{{\rm tr}}

\newcommand{\be}{\begin{equation*}}
\newcommand{\ee}{\end{equation*}}
\newcommand{\bea}{\begin{eqnarray*}}
\newcommand{\eea}{\end{eqnarray*}}

\newcommand{\avr}[1]{\langle#1\rangle}

\newcommand{\ra}{\rightarrow}
\newcommand{\HS}{{\mathfrak{H}}}

\newcommand{\x}{\rangle\langle}

\begin{document}

\newtheorem{theorem}{Theorem}
\newtheorem{lemma}[theorem]{Lemma}
\newtheorem{corollary}[theorem]{Corollary}
\newtheorem{proposition}[theorem]{Proposition}
\newtheorem{definition}[theorem]{Definition}
\newtheorem{example}[theorem]{Example}

\newenvironment{proof}{\vspace{1.5ex}\par\noindent\textbf{Proof }}%
    {\hspace*{\fill}$\Box$\vspace{1.5ex}\par}
\newenvironment{remark}{\vspace{1.5ex}\par\noindent{\it Remark}}%
    {\hspace*{\fill}$\Box$\vspace{1.5ex}\par}

\title{Assessing non-Markovian dynamics}

\author{M.M.\ Wolf$^{1,2}$,
J.\ Eisert$^{3,4}$,
T.S.\ Cubitt$^{5}$,
and J.I.\ Cirac$^{1}$}
\affiliation{$^{1}$Max-Planck-Institute for Quantum Optics, 
85748 Garching, Germany\\
 $^{2}$Niels Bohr Institute, University of Copenhagen, 17 2100 Copenhagen, Denmark,\\
 $^{3}$Institute of Physics and Astronomy, 
University of Potsdam, 14476 Potsdam,  Germany\\
 $^{4}$Quantum Optics and Laser Science, 
 Imperial College London, London SW7 2PE, UK\\
$^{5}$Department of Mathematics, University of Bristol, Bristol BS8 1TW, UK}

\date{\today}

\begin{abstract} We investigate what a snapshot of a quantum evolution---a quantum 
channel reflecting open system
dynamics---reveals about the underlying continuous time
evolution. Remarkably, from such a snapshot, and
without imposing additional assumptions,
it can be decided whether or not a channel is consistent with
a time (in)dependent Markovian evolution, for which we
provide computable necessary and sufficient criteria.
Based on these, a computable measure of
`Markovianity' is introduced. We discuss how the
consistency with Markovian dynamics can be checked
in quantum process tomography. The results also clarify
the geometry of the set of quantum
  channels with respect to being solutions of 
  time (in)dependent
  master equations.
\end{abstract}

\maketitle

\section{Introduction}

Much of the power of information theory, classical and quantum,
comes from the separation of information from its physical
carriers. This level of abstraction favors a black box approach
describing physical processes by their input-output relations in
discrete time steps. For quantum systems the most general black
box is described by a trace preserving and completely positive
map---a quantum channel. This might describe the application of a
gate in a quantum processor, a quantum storage device, a
communication channel, or any open systems dynamics where one
merely has partial access to the relevant degrees of freedom.
In any case it can be considered a snapshot of a physical evolution after a certain time.

In the present letter we investigate what such a snapshot reveals
about the intermediate continuous time evolution, in particular
regarding the Markovian and hence memoryless
or non-Markovian character of the process. This will link the
black box approach to the dynamical theory of open quantum
systems. Remarkably, fixing a single point in time
enables us to gain non-trivial information about the path along
which the system has or has not evolved, even without making
additional assumptions about the physics of the environment and
its coupling to the system. On the one hand this analysis thus
provides a model-independent means of investigating non-Markovian
features. On the other hand it tells us which type of evolution is
required for the continuous realization of theoretically given
quantum channels.

On the experimental side recent progress in the field of quantum
information science showed more and more precise determination of
input-output relations via \emph{quantum process tomography}.
This has by now been achieved in various systems including NMR
\cite{TeleNature}, ion traps \cite{Blatt}, linear
optics implementations \cite{OBrien} and solid state qubits
\cite{Howard}. Some of them, in fact, rely on
the a priori assumption that the process in fact is Markovian.
In the study of open quantum systems as such,
non-Markovian processes also move to the center
of interest \cite{Structured,Piilo}.
In general, a precise understanding of decoherence and
dissipation processes is vital for further improvements and for
the design of adapted error correcting codes and
fault-tolerant schemes. Most notably in this context
non-Markovian effects are known to require a careful
analysis \cite{Preskill}.

Before we start, some remarks concerning the central notions
`Markovian' and `time-dependent Markovian' are in order. We will
call a quantum channel \emph{Markovian} if it is an element of any
one-parameter continuous completely positive semigroup, i.e., a
solution of a master equation with generator in Lindblad form. If
the generator depends on time, we use the term
\emph{time-dependent Markovian} instead. In both cases the
continuous evolution is memoryless in that at any point
in time the future evolution only depends on the present state and
not on the history of the system. Our findings are:

\begin{itemize}
\item[(i)] The sets of (time-dependent) Markovian channels are
strictly included within the set of all quantum
channels and exhibit a non-convex geometry.

\item[(ii)] For arbitrary finite dimensions there is an  efficient
algorithm for deciding whether or not a quantum channel is
Markovian.

\item[(iii)] A computable measure is introduced which quantifies the
Markovian part of a channel.

\item[(iv)] For qubits a simple criterion
for time-dependent Markovianity
is given together with a detailed analysis of the geometry of the
sets of quantum channels.

\item[(v)] Examples of non-Markovian processes are discussed.

\item[(vi)] An application in the theory of renormalization group (RG)
transformations on quantum spin chains is outlined.
\end{itemize}

\section{Preliminaries}\label{sec:pre}
Throughout we will consider
quantum channels on finite dimensional systems, i.e., linear maps
$T:\M_d\ra\M_d$ on $d\times d$ (density) matrices,
$\rho\mapsto T(\rho)$
referred to as \emph{dynamical maps}, reflecting the snapshot
in time \cite{Uncor}. 
When occasionally changing from Schr\"odinger to
Heisenberg picture we
will denote the respective map by $T^*$. It will be convenient to
consider $\M_d$ as a Hilbert space $\HS$ equipped with the scalar
product $\avr{A,B}_{\HS}=\tr[A^\dagger B]$. On this space the map
$T$ is represented by a matrix 
\begin{equation*}
	\hat{T}_{\alpha,\beta}=
	\tr[F_\alpha^\dagger
	T(F_\beta)]=\avr{F_\alpha|T|F_\beta}_\HS, 
\end{equation*}	
where
$\{F_\alpha\}_{\alpha=1,\ldots, d^2}$ is any orthonormal basis in
$\HS$. Unless otherwise stated, we will use matrix units $\{|i\x
j|\}_{i,j=1,\ldots, d}$ as basis elements. Note that a concatenation
of two maps $T_1, T_2$ simply corresponds to a product of the
respective matrices $\hat{T_2}\hat{T_1}$ and that a density matrix
$\rho$ in this language becomes a vector with entries $\langle
i,j|\hat{\rho}\rangle=\langle i |\rho|j\rangle$. A useful
operation is the involution
\begin{equation*}	
	\avr{i,j|\hat{T}^\Gamma|k,l}=\avr{i,k|\hat{T}|j,l}
\end{equation*}	
\cite{involution,divisibility}. 
It connects the matrix representation $\hat{T}$
of the map to its Choi matrix 
\begin{equation*}
	\hat{T}^\Gamma=d
	(T\otimes\id)(\omega) 
\end{equation*}
where $\omega$ is a
maximally entangled state $\omega=|\omega\rangle
\langle\omega|$, 
$|\omega\rangle=\sum_{i=1}^d| i,i \rangle
/\sqrt{d}$. Complete positivity is then equivalent to
$\hat{T}^\Gamma\geq 0$. \emph{Quantum channels} are completely
positive and trace preserving maps.

 The workhorse in the dynamical theory of open
 quantum systems are semigroups $\{e^{t L}\} $
  depending continuously on one parameter $ t\geq 0$
 (time) and giving rise to completely positive evolution for all
 time intervals.
 Two equivalent standard forms for the respective generators have
 been derived in Refs.\ \cite{Lindblad}:
 \begin{equation}
 L(\rho) = i [\rho,H] + \sum_{\alpha,\beta}
 G_{\alpha,\beta} \bigl(F_\alpha\rho F_\beta^\dagger -
 \frac12\{F_\beta^\dagger
 F_\alpha,\rho\}_+\bigr), \label{normal2}
 \end{equation}
 so $L(\rho)= \phi(\rho) - \kappa\rho - \rho\kappa^\dagger$,
 where  $G\geq
 0$, $H=H^\dagger$, $\phi$ is completely positive and
 $\phi^*(\1)=\kappa+\kappa^\dagger$. A channel will be called
 \emph{time (in)dependent Markovian} if it is the solution of any  master equation
 $\dot{\rho}=L(\rho)$ with time
 (in)dependent Liouvillian in \emph{Lindblad form} (\ref{normal2}), 
i.e., 
\begin{equation*}
	T=\exp\left({\int_0^1dt\;
	 L_t}\right)\,\text{ time-ordered}.
\end{equation*}

\section{Deciding Markovianity}

Given a quantum channel $T$ when is it
Markovian, i.e, of the form $T=e^L$? A priori, this might be a
trivial question: as the channel fixes only one point within a
continuous evolution, there might always be a `Markovian path'
through that point. As the attentive reader might already guess,
this turns out to be not the case. One attempt to decide whether
$T$ is Markovian could be to start from a Markovian ansatz and
then calculate $\inf_L||T-e^L||$, e.g., by numerical minimization.
The major drawback of such an approach is the non-convex geometry
of the set of Markovian channels \cite{NonConvex}
which inevitably leads to the
occurrence of local minima. The following approach circumvents
this problem and guarantees to find the correct answer efficiently
by first taking the $\log$ of $T$ and then deciding whether this
is a valid Lindblad generator of the form (\ref{normal2}).

The latter can easily be decided: a map $L:{\cal
M}_d\rightarrow{\cal M}_d$ can be written in Lindblad form iff (a)
it is Hermitian \cite{Herm}, (b) $L^*(\1)=0$ corresponding to
the trace preserving property and (c) $L$ is \emph{conditionally
completely positive} (ccp) \cite{ccp}, i.e.,
\begin{equation}
	 \label{eq:ccp}
	\omega_\perp \hat{L}^\Gamma \omega_\perp\geq 0\;,
\end{equation}
where $\omega_\perp=\1-\omega$ is the projector onto the
orthogonal complement of the maximally
entangled state (see appendix).

Before applying (\ref{eq:ccp}) to $\log \hat T$ we need to
discuss some spectral properties of quantum channels.
 For simplicity we will restrict ourselves to the generic case
 where
$\hat{T}$ has non-defective and non-degenerate
Jordan normal form.
Hermiticity of a channel implies that its eigenvalues
are either real or come in complex conjugate pairs. The {\it Jordan
normal form}---achievable via a similarity transform
in the orthonormal basis in which the channel is expressed---is then
\begin{equation}
	\hat{T} =\sum_r \lambda_r P_r + \sum_c
	\lambda_c P_c + \bar{\lambda_c}
	\mathbb{F}\bar{P_c}\mathbb{F},\label{eq:Jordan}
\end{equation}
where $r$
labels the real and $c$ the complex eigenvalues respectively. The
$P$'s are orthogonal (but typically not self-adjoint) spectral
projectors and $\mathbb{F}$ is
 the
flip-operator ($\mathbb{F}|a\rangle\otimes|b\rangle=
	|b\rangle\otimes|a\rangle$).
Projectors corresponding to complex conjugate eigenvalues are
related via $P\leftrightarrow\mathbb{F}\bar{P}\mathbb{F}$  due to
Hermiticity of the channel which can in turn be expressed as
$\mathbb{F}\hat{T}\mathbb{F}=\bar{\hat{T}}$.

Now $T$ is Markovian iff there is a branch of the logarithm $\log
\hat{T}$, defined via the logarithm of the eigenvalues in
Eq.\ (\ref{eq:Jordan}), which fulfills the above mentioned
conditions (a) -- (c). Note that (b) is always satisfied if we
start from a trace preserving map. Moreover, Hermiticity holds iff
there is no negative real eigenvalue and branches for each
complex pair of eigenvalues are chosen consistently so
that the eigenvalues remain complex
conjugates of each other. If $\hat{T}$ has negative
eigenvalues, the dynamics will not be Markovian.
The set of Hermitian logarithms is then
characterized by a set of integers $m_c\in\mathbb{Z}$,
\begin{equation}
	\label{eq:logs} \hat{L}_m=\log{\hat T} = \hat{L}_0 + 2\pi
	i \sum_c m_c \big(P_c-\mathbb{F}\bar{P_c}\mathbb{F}\big),
\end{equation}
where $\hat{L}_0$ denotes the principal branch. The infinity of discrete
branches looks a bit awkward at first glance, but the problem can
now be cast into a familiar form. Defining the matrices 
\begin{eqnarray*}
	A_0&=& \omega_\perp\hat{L}^\Gamma_0\omega_\perp,\\
	A_c&=& 2\pi i\;
	\omega_\perp\big(P_c-\mathbb{F}\bar{P_c}\mathbb{F}\big)^\Gamma\omega_\perp
\end{eqnarray*}
and applying Eq.\ (\ref{eq:ccp}) to $\hat{L}_m$ yields that $T$ is Markovian iff
 \begin{equation}
	 \label{eq:semi0} A_0 +\sum_c m_c A_c
	\geq 0
\end{equation}
for any set of integers $\{m_c\}$.
These matrices $A_0,A_c$
will be Hermitian if $\hat{T}$ has only positive real eigenvalues.
Note that the real solutions $m_c\in\mathbb{R}$ of Eq.\
(\ref{eq:semi0}) form a convex set
${\cal S}\subseteq\mathbb{R}^C$ ($C$ being the number of
complex pairs of eigenvalues). The present problem is then to
decide whether $\cal S$ contains an integer point. Fortunately,
this has an efficient solution in terms of a semi-definite integer
program. That is, there is an algorithm \cite{sdpi} which either
finds a solution or guarantees that none exists within a run time
of order $l d^2$ where $l$ is the number of digits to which the
input is specified \cite{Feasible}.

In practice it turns out that checking the vicinity of the
principle branch ($m_c\in\{-1,0,1\}$) is typically
sufficient. For qubit channels the problem simplifies
further since it becomes (at most) one-dimensional as $C\leq
1$. Hence, one has merely to maximize the smallest eigenvalue (a
concave function) of Eq.\ (\ref{eq:semi0}) with respect to real $m$
and then check positivity for the two neighboring integers.
Note that again, this criterion 
does not necessarily reveal whether the dynamics was
truly Markovian, but whether it is consistent with it.

\begin{figure}[ttt]
\begin{center}
\epsfig{file=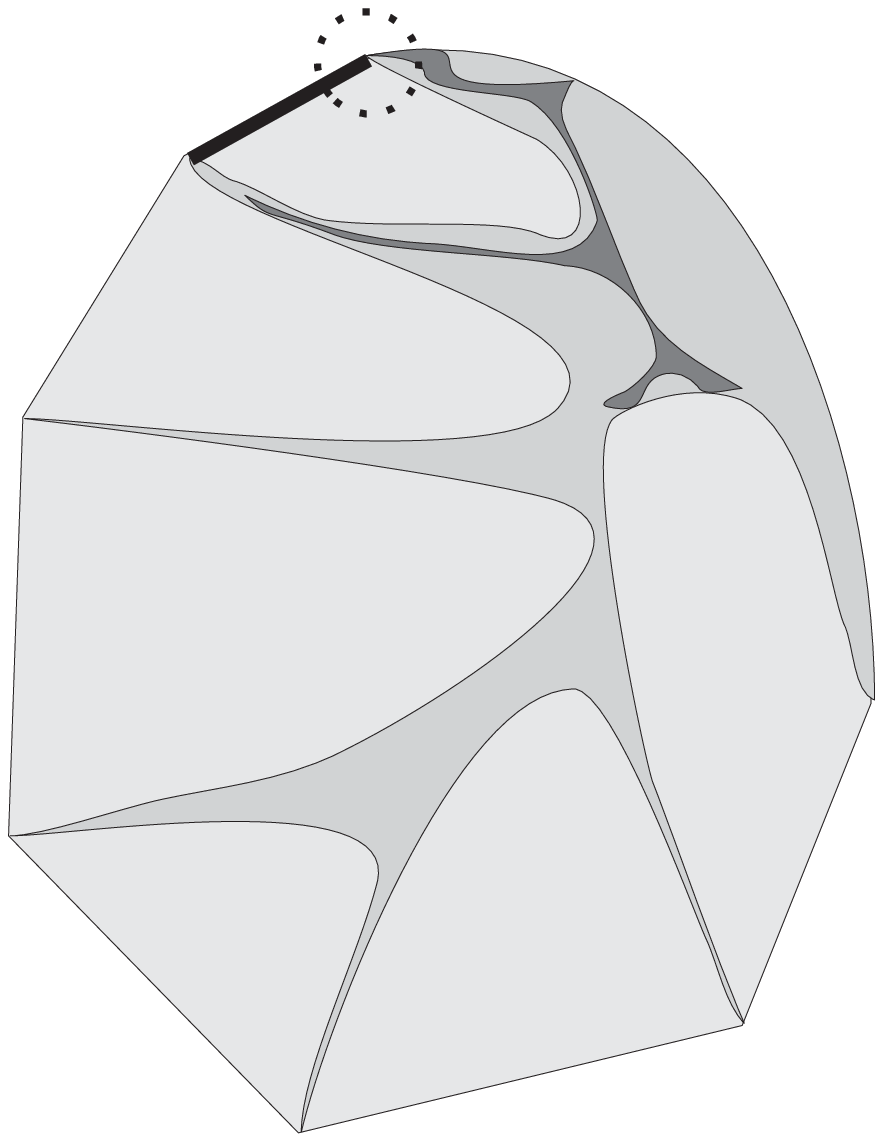,angle=-70,width=0.65\linewidth}
\vspace*{-0.4cm}
\end{center}
\caption{Schematic depiction of the (12-dimensional) convex set of
qubit channels. The (dark grey) subset of Markovian channels is
non-convex and contains 2\% of the channels. The larger still
non-convex set of time-dependent Markovian channels (17\%)
contains all extremal channels. All sets, including the measure
zero set of indivisible channels (black line) can be found in the
neighborhood of the identity (dotted circle).}
\label{figure:set}
\end{figure}

\section{Measuring Markovianity} 

When applying the above criterion to
random quantum channels one finds that only a small (but
remarkably nonzero) fraction of them is Markovian,
see Fig.\ \ref{figure:set}. In the following we
will show how one can quantify the deviation from
Markovianity for
the remaining channels. Desirable properties of a measure of
Markovianity $M$ are: 
\begin{itemize}
\item[(i)] some form of normalization, e.g.,
$M\in[0,1]$ with $M(T)=1$ iff $T$ is Markovian, 
\item[(ii)]
computability, 
\item[(iii)] continuity, 
\item[(iv)] basis independence, i.e.,
$M(U^\dagger T U)=M(T)$ for all unitary channels $U$ and 
\item[(v)] an
operational or physical interpretation.
\end{itemize}

One possibility would again be to start from a distance measure
$\inf_L||T-e^L||$ which, however, looses much of its appeal by the
apparent difficulties in computing it. 
We therefore propose a different approach
based on the criterion in Eq.\ (\ref{eq:semi0}). To this end let us
regard $L_m$ as Liouvillian of a master equation. If the channel
is not Markovian this will not give rise to completely positive,
i.e., physical, evolution for all times. However, adding an
additional dissipative term might yield a physical Markovian
evolution. If we choose isotropic noise of the form $
\rho\mapsto e^{-\mu}\rho + (1-e^{-\mu})\1/d,\quad\mu\geq
0$ 
with corresponding generator
$\hat{\cal L}_\mu=-\mu \omega_\perp$ then
$L_m+{\cal L}_\mu$ becomes a valid Lindblad
generator for some $m$
iff $\mu$ exceeds
\begin{equation}
	\mu_{\text{min}}=\inf\bigl\{\mu\geq 0:\exists
	m\in\mathbb{Z}^C:A_0+\sum_c m_cA_c+\frac{\mu}d\1\geq
	0\bigr\}.\nonumber
\end{equation}
Hence $\mu_{\text{min}}$ is the minimum
amount of isotropic noise required to make the channel
Markovian.
Note that $\mu_{\text{min}}$ can again be calculated by semidefinite
integer programming and that it is basis independent in the sense
of (iv). In order to meet the normalization condition and to add
an intuitive geometric interpretation to the physical one we use
\begin{equation*}
	M(T)=\exp\big[\mu_{\text{min}}(1-d^2)\big]\;\in[0,1]
\end{equation*}	
as a {\it measure for Markovianity}.  If $A_0$
is not Hermitian we assign $M(T)=0$.
This turns out to be precisely the factor by which
the additional dissipation shrinks the output space of the channel
in order to make it Markovian. In order to see this note that the
volume of the output space (one might think in terms of the Bloch
sphere for $d=2$) is quantified by the determinant of the channel
\cite{divisibility}. Moreover,
$\det(e^{L_m+{\cal
L}_{\mu_{\text{min}}}} )=e^{\tr{\hat{L}_m}}e^{\tr{\hat{\cal
L}_{\mu_{\text{min}}}}}=\det (T)M(T)$, since $\tr[\hat{L}_m]$ is
independent of $m$. In this sense $M(T)$ quantifies the Markovian
part of the channel \cite{MZ2}.

\section{Discussion}

\begin{figure}[ttt]
\begin{center}
\epsfig{file=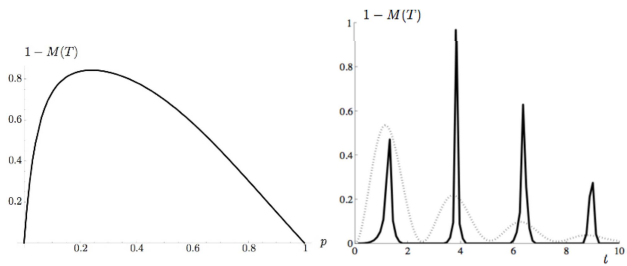,angle=0,width=.98\linewidth}
\vspace*{-0.4cm}
\end{center}
\caption{Deviation from Markovianity. (a) For a mixture of a
$\pi/4-\sigma_x$ rotation ($p=1$) and a dephasing channel
($p=0$). (b) For the
damped Jaynes Cummings model (as a function of time),
in which a single spin or qubit is coupled to a single cavity mode
undergoing lossy dynamics. The field mode serves as an
intermediate system preserving correlations that are relevant
for the systems's dynamics. The figure shows the interplay
between the time scale of truly irreversible cavity losses and
apparent decay on the time scale of oscillations, leaving
intervals which are consistent with a Markovian process
($\omega=0.2$, $\gamma=0.35$, $\alpha_{x,z}=1/2$,
$\alpha_y=
1$). Also shown (dotted) is the evolution of
$\langle 0|\rho|0\rangle$ for initial condition
$\langle 1|\rho|1\rangle=1$.}
\label{Fig:plot}
\end{figure}

Fig.\ \ref{Fig:plot} (a) shows the Markovianity of a convex combination
$T=p T_1+(1-p)T_2$ of a unitary channel $T_1$ corresponding to
$\pi/4$-Rabi oscillation (with Hamiltonian $\sigma_x$) and a
dephasing process $T_2=e^L$ with
\begin{equation*}
	L(\rho)=\sigma_z\rho\sigma_z-\rho. 
\end{equation*}	
This confirms the
{\it non-convex
geometry} in Fig.\ \ref{figure:set} and shows that non-Markovian
effects can arise from an environment which is in a mixture of
states each of which leads to a Markovian evolution.
Interestingly, there also exist non-Markovian processes that
could have arisen from a Markovian process when
judged from a snapshot in time: The {\it spin-star network}
in Ref.\ \cite{Structured} has the property that for all
times $A_0= \omega_\perp\hat{L}^\Gamma_0\omega_\perp
\geq0$, and hence the channel is consistent with
Markovianity. This is perfectly physical, as in each time step
there could have been a different memoryless evolution.
Fig.\ \ref{Fig:plot} (b) in turn
depicts the deviation from Markovianity for the {\it damped
Jaynes-Cummings model}, where the non-Markovian
character of the dynamics is clearly displayed \cite{JCH}.
Further examples
where this competing effect of time scales can be observed
are non-Markovian
models arising from spins coupled to {\it structured
baths} with an energy gap as studied, e.g., in Ref.\ \cite{Structured}.

\section{Time-dependent Markovian channels} 

For deciding whether a
channel is a solution of a time-dependent master equation 
we resort to generic qubit
channels \cite{DF} and content
ourselves with presenting results whose technical proofs are
published elsewhere \cite{divisibility}. A necessary and
sufficient criterion is most easily expressed
in the basis of Pauli matrices
$\{F_\alpha\}=\{\1,\sigma_x,\sigma_y,\sigma_z\}/\sqrt{2}$. Let
$g={\text{diag}(1,-1,-1,-1)}$ and $s_i$ be the ordered square
roots of the eigenvalues of $\hat{T}g\hat{T}g$. Then $T$ is
time-dependent Markovian iff $\det(T)> 0$ and
$ s_1^2s_4^2\geq
\prod_i s_i$ \cite{TD}. This set contains all extremal qubit channels
(Fig.\ \ref{figure:set}), but only 17\% of all
channels \cite{Percent}. A
paradigmatic example outside this set is
\begin{equation*}
	T(\rho)=
	\big(\tr[\rho]\1+\rho^T\big)/3,
\end{equation*}
which is the best physical
approximation to matrix transposition (or time-reversal
or---in optics---phase conjugation when
rotated by $\sigma_y$). This channels fails the criterion since
$\det(T)=-1/27$ \cite{divisibility}. In fact, it belongs to the
peculiar set of \emph{indivisible channels} that can not be
decomposed into a concatenation of two channels
unless one
of them is unitary.

\section{Addendum on quantum spin chains}

Let us finally outline an
application of the above results within an entirely different
field: RG transformations for translational invariant  states on
quantum spin chains. The transformation introduced in
Ref.\ \cite{RG}
amounts to a coarse graining consisting of two steps: (1.) 
build
equivalence classes of states which only differ by a change of
local basis, (2.) merge neighboring sites and then iterate. 
Both
steps can be carried out explicitly if the state's
 matrix product representation \cite{MPS} uses matrices of finite
 dimension $D$. Each equivalence class then
 corresponds to a quantum channel $T$ on ${\cal M}_D$ and merging
 neighboring sites is reflected by replacing $T$ with $T^2$. In
 this sense the renormalization group is identified with a semigroup of
 quantum channels.
 If $T$ is Markovian, we
can replace the discrete block spin transformation by an
 equivalent continuous RG flow 
 \begin{equation*}
 T\mapsto \exp[s \log T]
 \end{equation*}
  with $s\geq
 0$ parameterizing the scale of observation, see
 Eq.\  (\ref{eq:semi0}). 
 This allows then not only
 to coarse-grain ($s>1$) but also to zoom in
 ($s<1$).
Indivisible channels are also special in this context: Since
there is no $T_0$ such that $T_0^2=T$ they correspond to starting
points, i.e., ultraviolet limits, of RG flows. In fact, the best
physical approximation to time reversal in this way distinguishes the 
well known ground state of the spin-1 AKLT
Hamiltonian \cite{AKLT}.

\section{Summary}

In this letter, we have introduced a framework to
assess whether a given dynamical map describing a physical
process could have arisen from Markovian dynamics. To
test this property we have provided necessary and sufficient
conditions. We also introduced a natural measure of
Markovianity, quantifying the Markovian content of a process.
As such, we have provided the means to judge the
forgetfulness of a physical process from a snapshot in time.

\subsection*{Acknowledgments}

MW thanks G.\ Giedke and I.\
de Vega for valuable discussions. This work has been
supported by the Elitenetzwerk Bayern, the EU 
(QAP, COMPAS, Scala), the EPSRC, 
Microsoft Research, and the EURYI.

\section*{Appendix: Deciding Lindblad form}

The following  is a modification of the argument given in
Ref.\ \cite{ccp} and allows to decide whether or not a map $L$ can be
written in Lindblad form Eq.\ (\ref{normal2}). It is obvious from
(\ref{normal2}) that $L$ has to be Hermitian, i.e., $L(X)^\dagger=
L(X^\dagger)$ for all $X$, and that $L^*(\1)=0$. Necessity of Eq.\ (\ref{eq:ccp}) is seen
by exploiting Eq.\ (\ref{normal2}): 
\begin{equation*}
\hat{L}^\Gamma =
[ (\phi\otimes\id)(\omega) - (\kappa\otimes\1)\omega -
\omega(\kappa\otimes\1)^\dagger] d.
\end{equation*}
On the r.h.s. the first
term is positive and the
other terms vanish when projected onto $\omega_\perp=(\1-\omega)$.
Conversely any Hermitian matrix fulfilling  inequality
(\ref{eq:ccp}) is of the form $\hat{L}^\Gamma= P -
|\psi\rangle\langle\omega|-|\omega\rangle\langle\psi|$, with any
$P\geq 0$ and  $\psi\in\mathbb{C}^d$. To arrive at
Eq.\ (\ref{normal2}) we interpret $P$ as Choi-matrix of a completely
positive map $\phi$ and set
$(\kappa\otimes\1)|\omega\rangle=|\psi\rangle$.


\begin{thebibliography}{30}
\bibitem{Uncor}
	The definition of a general quantum channel
	always implicitly assumes that system and
	environment are initially uncorrelated.
\bibitem{NonConvex}
	Note that the closure of the set of Markovian 
	channels is also non-convex. Simple examples like 
	convex combinations of the identity channel and a
	flip of two levels can---depending on the 
	regime---be proven to be Markovian and 
	non-Markovian.
\bibitem{Herm}	
	A map on ${\cal M}_d$ is Hermitian if for
	all $X:\;L(X)^\dagger=L(X^\dagger)$ or equivalently
	$\hat{L}^\Gamma$ is Hermitian.
\bibitem{Feasible}
	For small dimensions $d$, this leads to a practical
	and efficient algorithm for testing for Markovianity of a quantum
	channel. Note that the computational hardness in the system's
	dimension $d$ is not due to any specific algorithm provided here,
	but in fact due to the nature of the problem as such. 
	For practical purposes, yet, one
	is typically faced with channels of a given
	dimension $d$, and not with a family of channels 
	of increasing $d$.
\bibitem{MZ2}	
	Note that this measure
	is merely continuous up to a set of measure zero: 
	The logarithm becomes discontinuous
	at zero so that $M(T)$ can become discontinuous at singular
	channels.  In the degenerate case a second set of
	discontinuities appears (again of measure zero): Namely if a
	Markovian channel with pairwise negative eigenvalues is
	perturbed such that the eigenvalues remain real but the
	degeneracy is
	removed.
\bibitem{JCH}
	In the damped Jaynes
	Cummings model (Fig.\ \ref{Fig:plot} (b)),
	the Hamiltonian is 
	\begin{eqnarray*}
	H&=& (\alpha_x \sigma_x + \alpha_y
	\sigma_y + \alpha_z \sigma_z)
	\otimes \ii + \ii\otimes a^\dagger a\\
	&+& \Omega
	\left(
	\sigma_+ \otimes a +
	\text{h.c.}\right) ,
	\end{eqnarray*}
	with $a$ being the annihilation operator of the field mode,
	undergoing Markovian damping with a Lindblad
	generator 
	\begin{equation*}
	D(\rho) = \gamma 
	(a\rho a^\dagger - a^\dagger a\rho/2
	- \rho a^\dagger a/2), 
	\end{equation*}
	and $\sigma_+= (\sigma_x+i \sigma_y)/2$
	(see, e.g., Ref.\ \cite{Piilo}). The non-Markovian dynamics is the
	reduced dynamics of the spin.
\bibitem{TD}
	Any qubit channel is a convex combination of time-dependent
	Markovian evolutions. For $d=3$ this is no longer true---a counterexample being $T(\rho)=(\1-\rho^T)/2$.
\end{thebibliography}
\end{document}